\documentclass[10pt]{iopart}
\usepackage{graphicx}
\usepackage{cite}
\begin{document}

\title[Event-chain Monte Carlo simulations of two-dimensional decagonal colloidal quasicrystals]{Event-chain Monte Carlo simulations of the liquid to solid transition of two-dimensional decagonal colloidal quasicrystals}

\author{M Martinsons, J Hielscher, S C Kapfer and M Schmiedeberg}
\address{Institut f\"ur Theoretische Physik I, Friedrich-Alexander-Universit\"at Erlangen-N\"urnberg, Staudtstra{\ss}e 7, 91058 Erlangen, Germany}
\eads{\mailto{miriam.martinsons@fau.de}, \mailto{johannes.hielscher@fau.de}, \mailto{sebastian.kapfer@fau.de}, \mailto{michael.schmiedeberg@fau.de}}	
\vspace{10pt}
\begin{indented}
\item[]June 2019
\end{indented}

\begin{abstract}
In event-chain Monte Carlo simulations we model colloidal particles in two dimensions that interact according to an isotropic short-ranged pair potential which supports the two typical length scales present in decagonal quasicrystals. We investigate the assembled structures as we vary the density and temperature. Our special interest is related to the transition from quasicrystal to liquid. We find a one-step first-order melting transition without a pentahedratic phase as predicted within the KTHNY melting theory for quasicrystals. However, we discover that the slow relaxation of phasonic flips, i.e.\@ rearrangements of the particles due to additional degrees of freedom in quasicrystals, changes the positional correlation functions such that structures with long-range orientational correlations but exponentially decaying positional correlations are observed.
\end{abstract}


\vspace{2pc}

\maketitle
\ioptwocol
\section{\label{sec:introduction}Introduction}
Melting in two dimensions is an interesting and intriguing process. While in three dimensions the phase transition between a solid crystal and a liquid is of first order \cite{Hoover}, in two dimensions there exist different competing melting theories. Solids in two dimensions possess long-range orientational order and quasi-long-range positional order \cite{Mermin,Mermin2}. The KTHNY theory \cite{Kosterlitz,Halperin,Young,Nelson} which was developed by Kosterlitz and Thouless \cite{Kosterlitz} (and independently by Berezinskii \cite{Berezinskii}) and completed by Halperin, Nelson and Young \cite{Halperin,Young,Nelson} predicts the melting to occur in two steps via an intermediate hexatic phase between solid and liquid. In the hexatic phase free dislocations cause short-range positional and quasi-long-range orientational order, expressed by exponentially and algebraically decaying correlation functions respectively. The solid-hexatic and the hexatic-liquid transition are both predicted to be continuous. The two-step melting is confirmed in numerous simulations and experiments \cite{Murray,Zahn,Keim,Lin,Lee,Shiba,Qi2,Prestipino}. However, systems of hard disks suggest a hexatic-liquid transition of first order \cite{Bernard,Engel3,Qi,Thorneywork}. In case of soft disks the hexatic-liquid transition depends on the parameters of the system. Increasing softness in purely repulsive potentials leads to a change from first-order to continuous transitions \cite{Kapfer}. Similar observations are made for further model systems with rising densities \cite{Zu}. Alternative theories (see, e.g. \cite{Strandburg,Alba-Simionesco}) predict a first-order solid-liquid transition which is either induced by grain boundaries \cite{Fisher,Chui,Chui2} or due to the simultaneous dissociation of dislocations and disclinations as studied in mean-field theories \cite{Kleinert}. Another theory suggests a melting induced by vacancies \cite{Joos}.

The melting of quasicrystals, i.e.\@ structures with long-range orientational order but no periodic translational symmetry \cite{Shechtman,Levine}, has hardly been studied yet. A melting theory for two-dimensional pentagonal quasicrystals was developed by De and Pelcovits \cite{De2,De3,De4,De} who predicted the melting to occur in two steps with an intermediate phase called pentahedratic phase in analogy to the hexatic phase in crystals \cite{De}. The process is supposed to be similar to the melting of periodic crystals as predicted by the KTHNY theory.

A characteristic of quasicrystals is their additional degrees of freedom that lead to phasonic excitations \cite{Levine2,Socolar} and cause complex rearrangements of the particles \cite{Kromer,Sandbrink,Martinsons,Hielscher}. Single local rearrangements are referred to as phasonic flips. The correlation of phasonic displacements $\varphi$ is expected to diverge as $\langle(\varphi(0) - \varphi(R))^{2}\rangle \propto \textrm{ln}(R)$ with distance $R$ in two-dimensional quasicrystals \cite{Kalugin,Henley}, i.e.\@ perfect long-range phasonic order is not possible. 

A suitable method to sample equilibrium configurations of large systems efficiently is event-chain Monte Carlo simulations which were introduced by Bernard $et \, al.$ \cite{Bernard2} and initially applied to investigate systems of hard particles \cite{Bernard2,Bernard,Kampmann,Kampmann2,Isobe}. Recently, the algorithm was generalized to particles that interact according to continuous pair potentials \cite{Michel} and applied to purely repulsive pair potentials \cite{Kapfer}. The event-chain Monte Carlo algorithm is rejection-free and known to be faster than conventional Monte Carlo algorithms \cite{Michel}. 

In this article we extend the event-chain Monten Carlo algorithm to a pair potential of the Lennard-Jones--Gauss type \cite{Engel1,Engel2} with two minima that support the two incommensurate length scales present in decagonal quasicrystals. We model a phase diagram dependent on temperature and density. We mainly focus on the transition from quasicrystal to liquid and discuss the influence of phasonic degrees of freedom on the order of the structures.

The article is organized as follows: In section \ref{sec:methods} we describe our system as well as the simulation method. We present and discuss the results of our simulations in section \ref{sec:results} and give a conclusion in section \ref{sec:conclusions}.

\section{\label{sec:methods}Methods and system}

\subsection{\label{sec:meth} System}

We simulate two-dimensional systems of identical particles that interact according to an isotropic Lennard-Jones--Gauss pair potential \cite{Engel1,Engel2} 
\begin{equation}
\frac{V(r)}{\epsilon_{0}} = \left(\frac{r_{0}}{r}\right)^{12} - 2\left(\frac{r_{0}}{r}\right)^{6} - \epsilon \, \textrm{exp} \left(- \frac{(r - r_{G})^{2}}{2 r_{0}^{2} \sigma^{2}} \right).
\end{equation}

The additional Gaussian term causes a second minimum in the potential. $\epsilon_{0}$ sets the energy unit, $\epsilon_{0}/k_{B}$ the temperature unit, and $r_{0}$ the length unit. The number density $\rho=N/(L_{x} L_{y})$ where $N$ is the number of particles and $L_{x}$ and $L_{y}$ denote the box lengths is given in units of $1/r_{0}^{2}$. In accordance with \cite{Engel1} we choose the potential parameters $r_{G}=1.52$, $\epsilon = 1.8$ and $\sigma^{2}=0.02$ such that two incommensurate length scales that are typical of decagonal quasicrystals are supported (see figure \ref{fig:overview} (a)). We truncate the interactions between the particles at a cutoff distance $r_{\mathrm{cut}}=2.5$, which is larger than the second minimum, and subtract the potential value at the cutoff to ensure continuity. 

\subsection{\label{sec:meth_1}Event-chain Monte Carlo simulation}

The applied event-chain Monte Carlo algorithm is based on chains of displacements as previously described in \cite{Bernard2}: A randomly chosen particle is displaced in successive infinitesimal steps in a given direction. Once a collision -- called event -- with another particle occurs, the displacement is transferred to the collision partner, which as a consequence is displaced in the same direction until another collision occurs. When the chain of displacements reaches a chosen displacement length, the procedure is continued with a new random initial particle, a new displacement direction, and a new displacement length. In case of continuous potentials \cite{Michel,Kapfer}, a collision event occurs if the potential barrier between two particles exceeds a given value. The potential barrier is given by the pair energy between the two particles. The allowed rise of the pair energy is determined randomly and individually for each particle pair. The cumulative number of events per particle is given in sweeps, i.e.\@ 1 sweep corresponds to $N$ events. 

We further extend the algorithm to our Lennard-Jones--Gauss potential with two minima, i.e.\@ we model two repulsive and two attractive regions. We only search for short-range events with possible event partners within the cutoff radius. 

In order to obtain a phase diagram and to investigate phase transitions, we model systems with various densities $\rho$ and temperatures $T$ in the $NVT$-ensemble. The simulation box is chosen with periodic boundary conditions and with a size that makes a periodic approximant structure that is similar to the perfect decagonal quasicrystal fit into the box. In our examples we either simulate $N=645$ particles in a box with $L_{x}/L_{y}\approx 1.91030$ or $N=18698$ particles with $L_{x}/L_{y}\approx 1.17557$. Our simulations are either started from a liquid or a decagonal approximant. The positions of the perfect decagonal approximant are obtained from the most pronounced maxima of the interference pattern of five laser beams \cite{Mikhael,Mikhael2,Gorkhali,Schmiedeberg2,Schmiedeberg}.

\subsection{\label{sec:meth_2}Analysis tools}

We make use of several analysis tools to identify the order of the structures. We analyze the rotational symmetry by means of the structure factor
\begin{eqnarray}
S(\bi{q}) \propto \sum_{j=1}^{N} \sum_{k=1}^{N} \mathrm{exp} \left[ 2\pi i \bi{q}(\bi{r}_k - \bi{r}_j)\right]
\end{eqnarray}
in reciprocal space. $\bi{q}$ denotes the wave vector and $\bi{r}_j$ the position of particle $j$. 

In order to characterize the structures with respect to 10-fold rotational symmetry we calculate the local bond-orientational order parameter \cite{Nelson2}
\begin{eqnarray}
\psi_{10} (\bi{r}_{j}) = 1/N_{k} \sum_{k=1}^{N_{k}} \textrm{exp}(10 \textrm{i} \theta_{jk}) 
\end{eqnarray}
for a particle $j$ at position $\bi{r}_{j}$. $\theta_{jk}$ denotes the angle between the bond from particle $j$ to a neighboring particle $k=1,...N_{k}$ and a fixed reference axis. We consider particles with a distance $d < (d_2 + d_3)/2$ as neighboring particles. $d_2$ and $d_3$ denote the second and third peak of the radial distribution function.

Furthermore, we consult the real-space tilings which are drawn by connecting nearest neighbor particles with a distance $d < (d_{1} + d_{2})/2$, where $d_{1}$ and $d_{2}$ denote the first and second peak of the radial distribution function.

\section{\label{sec:results}Results and Discussion}

In this section we present the results of our event-chain Monte Carlo simulations with  the pair potential and setup as described in the previous section. First, we give an overview of phases that form at various densities and temperatures. Afterwards, we study the phase transition between quasicrystal and liquid at a fixed temperature in detail. Third, we investigate the orientational and positional order of the structures by employing appropriate correlation functions. Finally, we discuss the impact of phason relaxation by means of correlation functions.

\subsection{\label{sec:phases} Phase behavior -- overview}

\begin{figure*}[htb]
\centering
\includegraphics[scale=1.1]{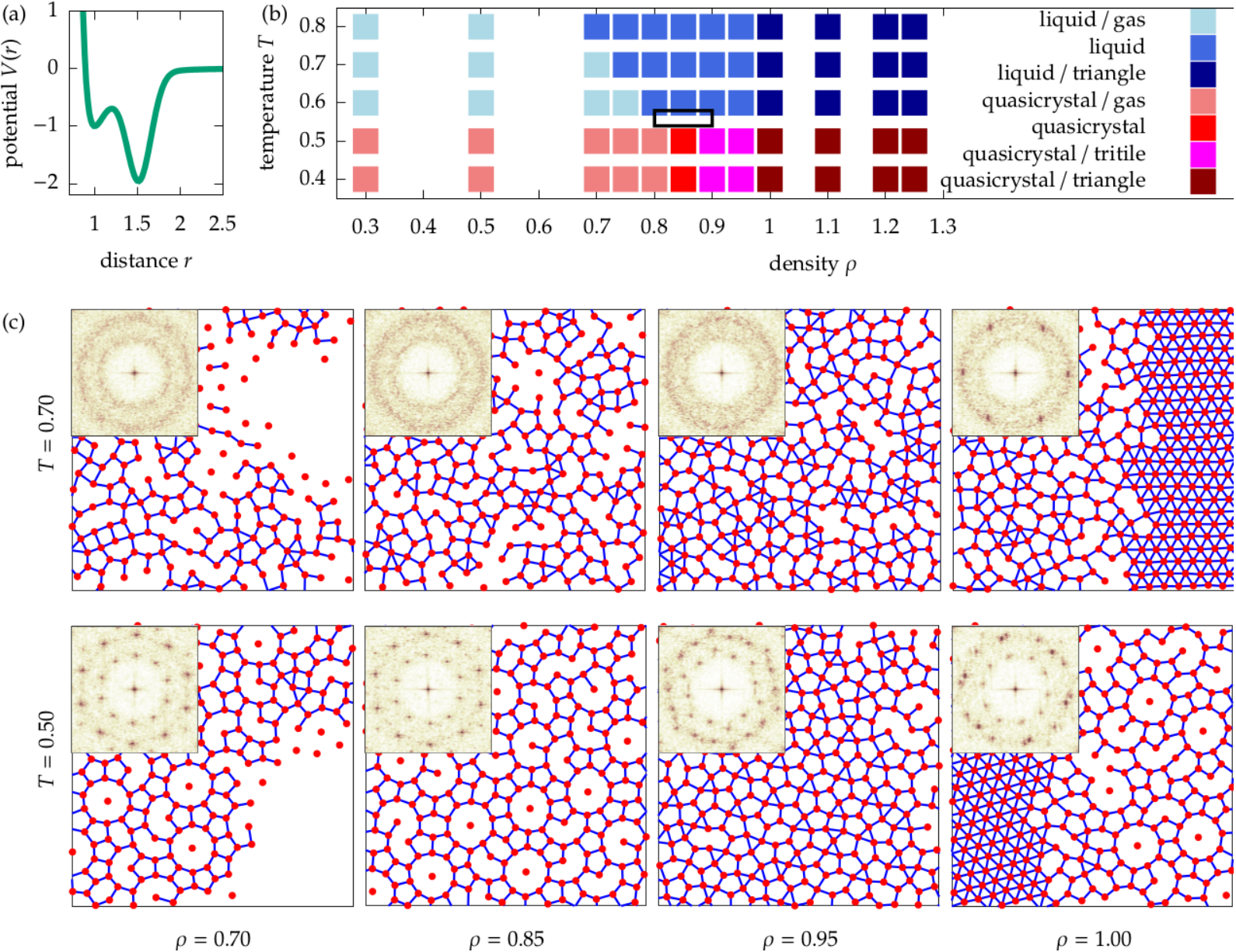} 
\caption{(a) Lennard-Jones--Gauss potential supporting the two incommensurate length scales present in decagonal quasicrystals. The potential parameters read $r_{G}=1.52,\, \epsilon = 1.8$ and $\sigma^{2}=0.02$. (b) Phase diagram dependent on the density $\rho$ and temperature $T$ in systems of $N=645$ particles. Simulations are started from the liquid. The color code on the right denotes the phases, which are either pure or a mixture of two coexisting phases marked by "/". The region within the black rectangle contains phase transitions from quasicrystal to liquid and will be investigated in detail in the next section. (c) Arrangements of particles at $T=0.70$ (upper row) and $T=0.50$ (lower row) and various densities $\rho=0.70$ (first column), $\rho=0.85$ (second column), $\rho=0.95$ (third column) and $\rho=1.00$ (fourth column). Nearest neighbor particles are connected. The insets depict the according structure factors.}
\label{fig:overview}
\end{figure*}

First, we investigate the phases that emerge in two-dimensional systems of $N=645$ particles interacting according to the Lennard-Jones--Gauss potential. The simulations are started from a liquid. The energies of the simulated systems fluctuate around constant energy values after about $3 \cdot 10^{6}$ or less sweeps. In order to check for any further relaxation we run the simulations for at least $5 \cdot 10^{6}$ sweeps. In the region where quasicrystals build we perform up to $8 \cdot 10^{7}$ sweeps. Dependent on the temperature and density we observe a rich variety of different structures which are summarized in a phase diagram (see figure \ref{fig:overview} (b)). We identify the phases by means of the structure factor and regarding the tilings as described in section \ref{sec:meth_2}.

For temperatures $T \geq 0.6$ the thermal energy of the particles is too large for the formation of quasicrystalline structures. Exemplary snapshots of systems with $T=0.7$ and various densities are visualized in figure \ref{fig:overview} (c) (upper row). We observe a coexistence between liquid and gas at low densities. The gas phase is dilute. With rising density the gas phase vanishes and the fraction of triangles with side lengths of the short potential length grows. At $\rho \geq 1.00$ the triangles arrange to an ordered lattice in coexistence with the liquid. Six peaks in the structure factor indicate hexagonal symmetry.

For low temperatures $T < 0.6$ the particles arrange to quasicrystalline structures at appropriate densities. Later in this article, we will have a closer look at the region where a phase transition from quasicrystal to liquid can be observed. The region is marked by a black rectangle in figure \ref{fig:overview} (b). In figure \ref{fig:overview} (c) (lower row) we illustrate snapshots with $T=0.5$ and various densities. At low densities a quasicrystal coexists with a gas. At $\rho \approx 0.85$ we get a nearly perfect quasicrystal with a structure factor with sharp peaks underlining the decagonal symmetry with globally constant orientation. Remarkably, at rising densities $\rho > 0.85$ the quasicrystal partially turns into a crystalline structure built from pentagons and triangles (see also \cite{EngelPhD}), which coexist with the quasicrystal. Note that pentagons belong to the five basic tiles of a decagonal quasicrystal and possess the highest particle densities among these. Triangles are supported by the short potential length. We refer to the arrangement as tritile phase. Its structure factor possesses a six-fold symmetry. Surprisingly, at densities that exceed the density of the tritile phase, the particles again rearrange to a quasicrystalline structure that now coexists with a hexagonal lattice with a lattice parameter of the short potential length. The structure factor shows a superposition of peaks that are arranged with decagonal and hexagonal symmetry respectively. The hexagonal region grows with rising density.
 
\subsection{\label{sec:transitions} Phase transitions}

\begin{figure*}[htb]
\centering
\includegraphics[scale=1.1]{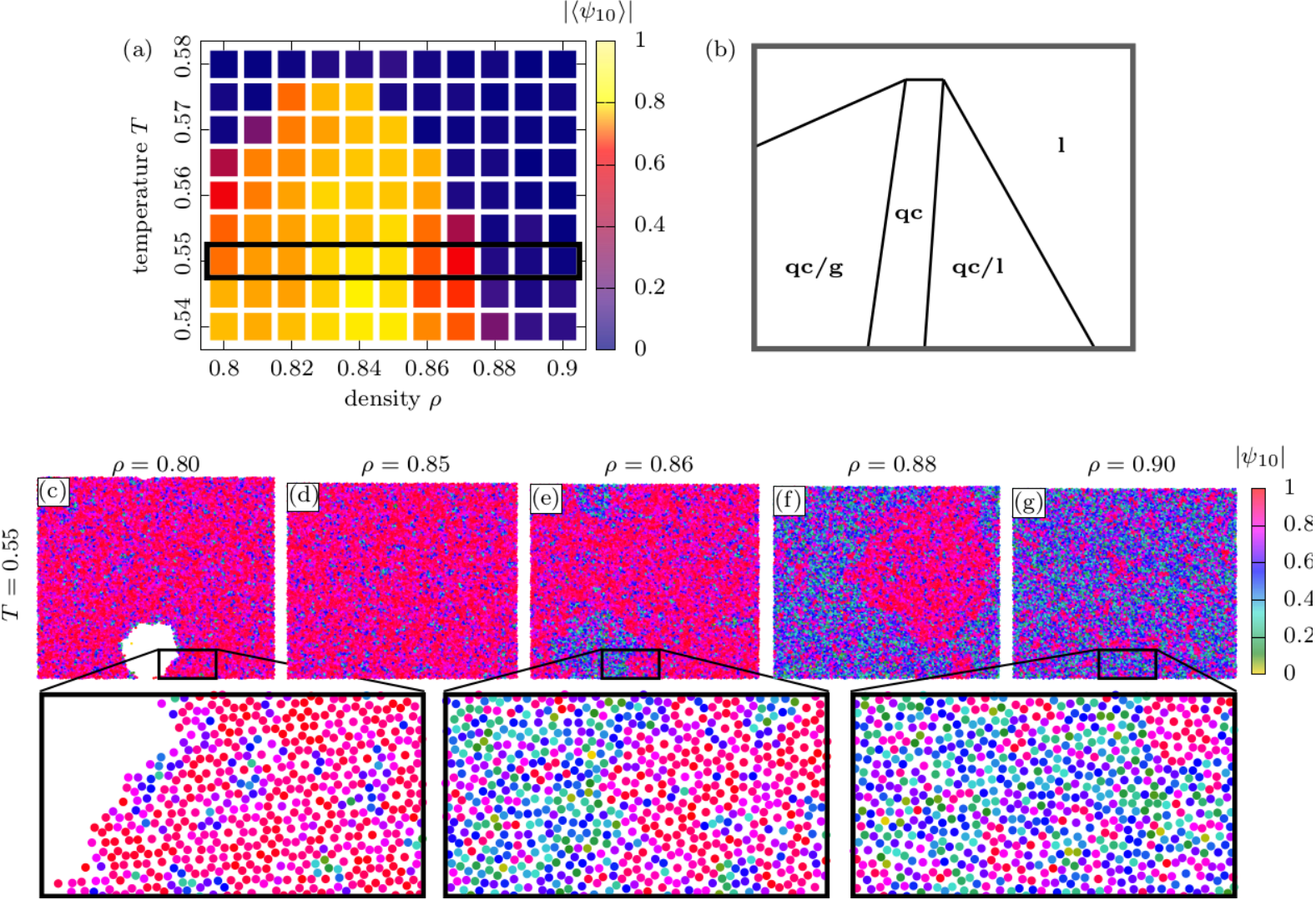}
\caption{Detailed view at the solid-liquid phase transition. (a) Absolute value of the average bond-orientational order parameter $|\langle \psi_{10} \rangle|$ in systems of $N=645$ particles with various densities and temperatures. A color code bar is on the right. The black rectangle indicates systems at a fixed temperature $T=0.55$ which are investigated in detail. (b) Rough schematic characterization of the corresponding phases consisting of quasicrystals (qc), gas (g) and liquid (l) or coexistences between two phases. (c)-(g) Arrangements of $N=18698$ particles at $T=0.55$ and various densities. The particles are colored according to their absolute value of the local bond-orientational order parameter $|\psi_{10}|$. The color code is shown on the right. (c) Coexistence between quasicrystal and gas at low density $\rho=0.80$. (d) Quasicrystal at $\rho=0.85$. (e) and (f) Coexistences between quasicrystal and liquid at $\rho=0.86$ and $0.88$. (g) Liquid at $\rho=0.90$. Bottom: Amplified particle-resolved sections of the systems. All depicted systems were simulated from the liquid.}
\label{fig.1}
\end{figure*}

In the following the region of the phase diagram in which we find quasicrystals is investigated in detail, i.e.\@ we present studies of phases with $\rho \in [0.8; 0.9]$ and $T \in [0.54; 0.58]$ that are received from simulations started with a liquid (see box in figure \ref{fig:overview}). In particular, we are interested in the transition from quasicrystal to liquid. In figure \ref{fig.1} (a) we plot $|\langle \psi_{10} \rangle|$ obtained from systems of $N=645$ particles as a function of the temperature and density. In figure \ref{fig.1} (b) a corresponding phase diagram is shown schematically. For sufficiently small temperatures the region with clearly largest $|\langle \psi_{10} \rangle|$ provides quasicrystalline order. The quasicrystal is surrounded by coexistences between quasicrystal and gas or respectively liquid at lower or higher densities.

We now choose a fixed temperature $T=0.55$ at which a solid-liquid phase transition is observed (see black rectangle in figure \ref{fig.1} (a)). We perform extensive simulations of large systems with $N=18698$ particles and various densities. In these large systems we perform at least $6 \cdot 10^{6}$ sweeps. In the region where quasicrystals build we even perform up to $2 \cdot 10^{7}$ sweeps. Snapshots of different phases obtained from simulations started with a liquid are visualized in figures \ref{fig.1} (c)-(g). The particles are colored according to $|\psi_{10}|$. 

The coexistence between quasicrystal and gas at low densities is accompanied with a sharp surface between both phases (see figure \ref{fig.1} (c)). At $\rho =0.85$ we observe a quasicrystalline phase with decagonal symmetry as illustrated in figure \ref{fig.1} (d). Small deviations from the decagonal symmetry in terms of phononic displacements are caused by thermal fluctuations. At $\rho \geq 0.86$ a liquid coexists with a phase with at least local quasicrystalline order (see figures \ref{fig.1} (e) and (f)). The phase coexistence suggests a first-order transition to the liquid. In principle, a similar behavior was observed for the phase transitions of periodic crystals \cite{Bernard,Kapfer}, where the coexistence is between a hexatic phase with local crystalline order and unbound dislocations and a liquid. However, in the next section we show that in our case the phase is not a pentahedratic phase (corresponding to the hexatic phase in periodic systems) as predicted in \cite{De}, but a quasicrystal. At $\rho \geq 0.90$ the orientational order is lost, resulting in a liquid (see figure \ref{fig.1} (g)). 

\begin{figure}[htb]
\centering
\includegraphics[scale=1]{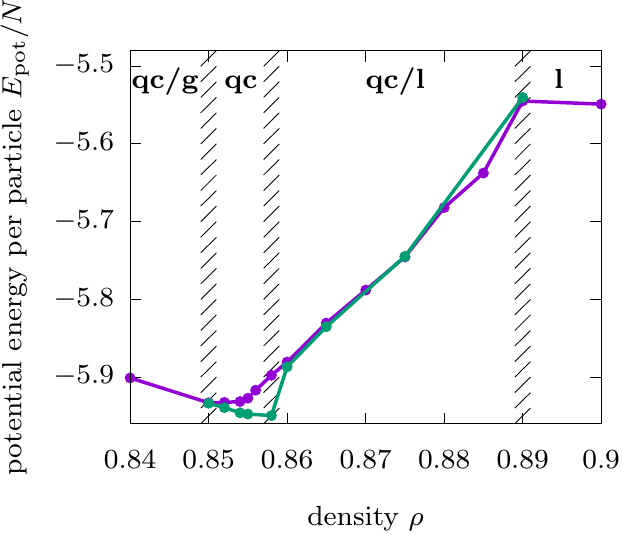} 
\caption{Potential energy per particle $E_{\mathrm{pot}}/N$ as a function of the density. The systems are simulated with $T=0.55$ and $N=18698$. The initial configurations are either a perfect decagonal approximant (green) or a liquid (purple). The corresponding phases are indicated on the top using the same abbreviations as in figure \ref{fig.1} (b). Lines roughly separate the phases.}
\label{fig:energy}
\end{figure}

In figure \ref{fig:energy} we illustrate the potential energy per particle $E_{\mathrm{pot}}/N$ of the investigated systems with $T=0.55$ and $N=18698$ as a function of the density $\rho$. Results are shown from simulations started with a liquid or a perfect decagonal approximant. The suggested first-order transition from the ordered phase to the liquid is supported. In case of simulations started from the solid the energy increases at slightly larger density $\rho \approx 0.86$ where coexistence with the liquid begins. Note that at low densities $\rho < 0.85$ the energy increases due to the coexistence with gas.

\subsection{\label{sec:correlation} Correlation functions}

\begin{figure}[htb]
\includegraphics[scale=0.9]{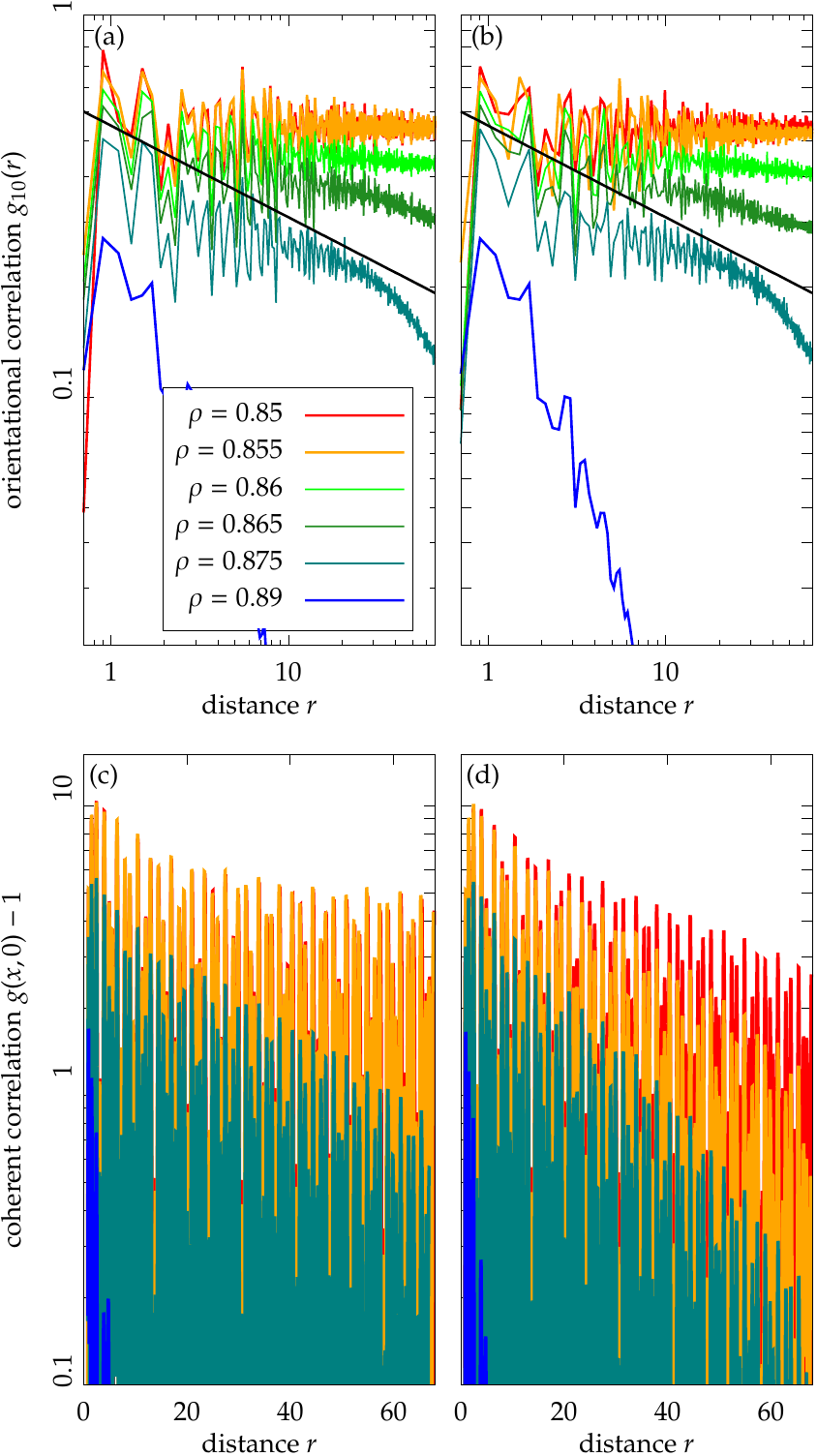}
\caption{(a)-(d) Correlation functions of systems with $T=0.55$ and various densities obtained from simulations started with a perfect decagonal structure (left column) or with a liquid (right column). (a) and (b) Bond-orientational correlation functions $g_{10}(r)$. The black lines illustrate the decay as predicted within the KTHNY theory for quasicrystals \cite{De} at the transition from pentahedratic to liquid, i.e.\@ $g_{10}(r) \propto r^{-0.25}$. Note that the correlation functions depicted in green are obtained from systems in the coexistence region. (c) and (d) Coherent correlation functions $g(x,0)$ along the symmetry directions.}
\label{fig.3}
\end{figure}

In order to decide whether the phase that coexists with the liquid is a solid quasicrystalline phase or a pentahedratic phase, we analyze the orientational and positional correlation functions of our systems with $N=18698$ and $T=0.55$ for various densities (see figure \ref{fig.3}). 

The orientational correlation function with respect to decagonal symmetry reads \cite{Nelson2,Zahn}
\begin{eqnarray}
g_{10}(r) = \langle \psi_{10}(\bi{r}_{j})\, \psi_{10}^{*}(\bi{r}_{k}) \rangle.
\end{eqnarray}
It describes the decay of the orientational order with the pair distance $r = |\bi{r}_{k} - \bi{r}_{j}|$. The star denotes the complex conjugate. For densities $\rho=0.85$ and $\rho=0.855$ below the beginning of the coexistence region, $g_{10}(r)$ is nearly constant which indicates long-range orientational order. This behavior is found in both, simulations started from an initially perfect quasicrystal (see figure \ref{fig.3} (a)), and from a liquid (see figure \ref{fig.3} (b)). Even at the beginning of the coexistence interval with densities $\rho=0.86$ and $\rho=0.865$, the correlations decay slower than the algebraic decay $g_{10}(r) \propto r^{-0.25}$ that is predicted to arise at a possible transition from a pentahedratic phase to the liquid \cite{De}. Accordingly, the orientational order suggests that there is no pentahedratic phase neither for low densities before the coexistence nor in coexistence with a liquid. In the liquid, the correlation functions decay exponentially and orientational order is lost. 

In periodic solids, the positional order is measured by means of the two-dimensional pair correlation $g(\bi{r})$. Here, we apply the method to our quasicrystals. In figures \ref{fig.3} (c) and (d) we depict the cut of $g(\bi{r})-1$ along the $x$-axis, i.e.\@ $\bi{r} = (x,0)$ (see also \cite{Bernard,Kapfer} for the so-called coherent pair correlation function). Simulations started from a perfect decagonal structure (see figure \ref{fig.3} (c)) show an algebraic decay of the peak height envelope for systems with low densities $\rho = 0.85$ and $\rho=0.855$. Such an algebraic decay is expected from a two-dimensional solid \cite{Mermin} and is in accordance with the results of the orientational order. In case of simulations started from a liquid (see figure \ref{fig.3} (d)) the positional correlation decays exponentially already for low densities $\rho=0.85$ and $\rho=0.855$, which indicates short-range positional order. Note that in periodic crystals, short-range positional order is not compatible with long-range orientational order. In a solid, the positional order should only decay algebraically as predicted by the Mermin-Wagner theorem \cite{Mermin,Mermin2}. 

\subsection{\label{sec:phason} Dislocations and phason relaxation}

In order to approach the origin of the short-range positional order, we analyze the considered structures regarding dislocations by filtering single density modes as previously described by Korkidi $et \, al.$ \cite{Korkidi}. In the quasicrystalline phase, both in the pure case and in coexistence with a liquid, we find a few pairs of bound dislocations as expected in a solid, no matter whether the system has been equilibrated from a perfect decagonal structure or from a liquid. However, we do not observe unbound dislocations, which are the characteristic features of the pentahedratic phase. This supports the absence of an intermediate pentahedratic phase during the phase transition, and the observed fast decay of the positional correlation function cannot be explained by dislocations. 

\begin{figure*}[htb]
\centering
\includegraphics[scale=0.9]{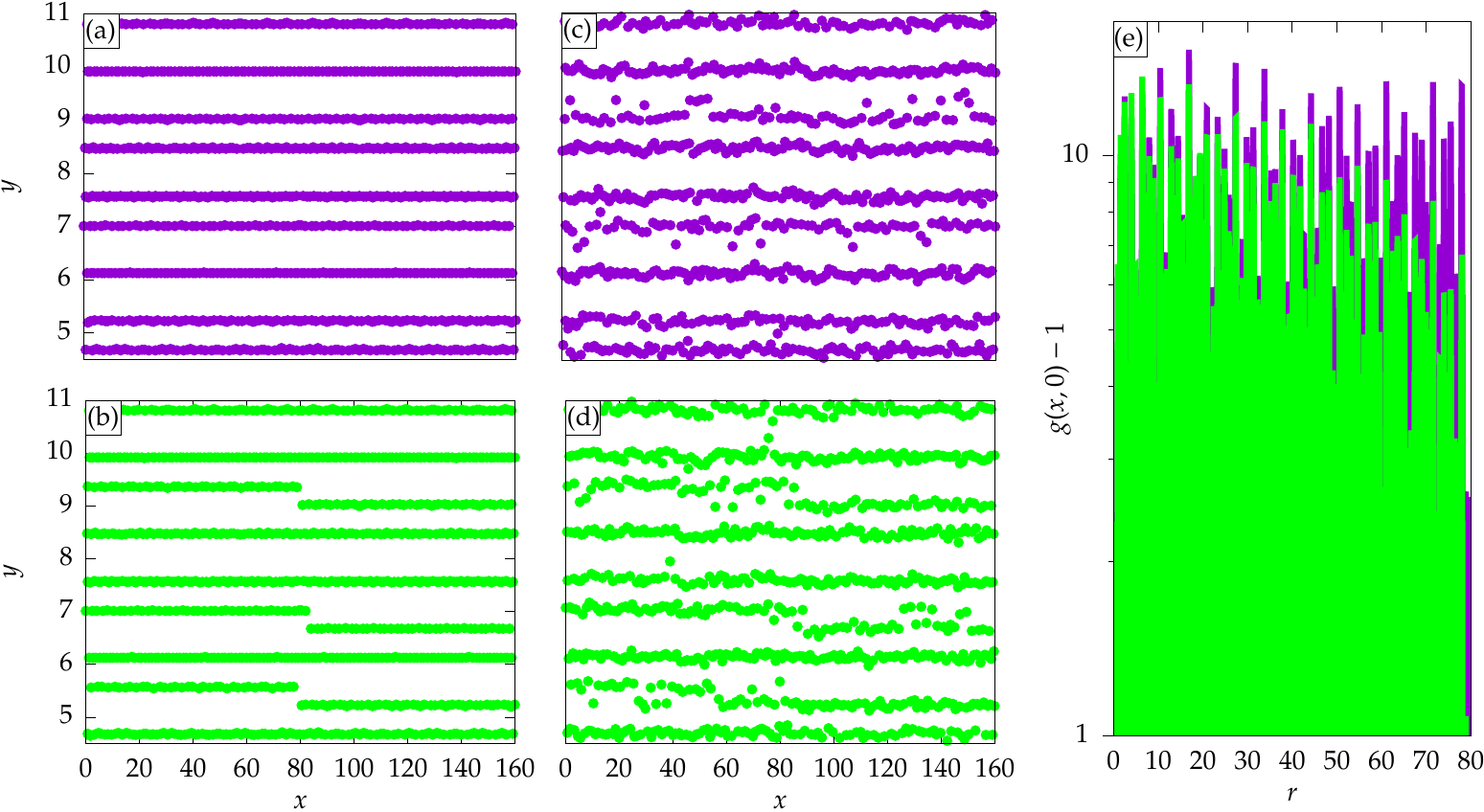} 
\caption{Oblique views on systems with decagonal order. The particle positions are shown in physical $(x,y)$ space. (a) Perfect decagonal approximant. (b) Structure from (a) with inserted correlated phasonic flips in the form of shifted lattice lines. (c) and (d) Structures from (a) and (b) heated at $T=0.5$. (e) Coherent correlation functions along the $x$-direction of the structures shown in (c) and (d).}
\label{fig:Phasons}
\end{figure*}

In a further approach we investigate how phasonic excitations impact the positional correlation function. Phasonic flips are local rearrangements that retain local decagonal symmetry, hence are energetically cheap. In a perfect decagonal structure, the positions of the particles are on well-positioned lines that are arranged according to a Fibonacci sequence (see figure \ref{fig:Phasons} (a)). Single flips cause only local deviations from the Fibonacci sequence and always retain quasicrystalline order. Furthermore, the flips can occur in a correlated way, and in the limit of excitations with infinite wave length, they do not cost any free energy at all. If phasonic excitations occur at a finite wavelength, incomplete lattice lines occur that destroy the global order according to the Fibonacci chain. We systematically insert correlated phasonic flips into the perfect decagonal structure by shifting half lattice lines (see figure \ref{fig:Phasons} (b)). The resulting structure is again a quasicrystal with neither local tiling defects nor dislocations. In Monte Carlo simulations we heat both systems at $T=0.5$ (see figures \ref{fig:Phasons} (c) and (d)). During heating, local flips occur, but the interrupted lines in the prepared system usually remain. Note that the systems are not meant to fully equilibrate, but serve as test systems for the decay of $g(\bi{r})$ in the presence of phasonic flips. The positional correlation functions along the $x$-direction of both heated systems are illustrated in figure \ref{fig:Phasons} (e). In case of the system with inserted phasonic flips the correlation function decays faster. Since no defects -- in particular no unbound dislocations -- are present, the fast decay can be attributed to the correlated flips. Consequently, $g(\bi{r})$ is not a clear criterion in order to determine the positional order in quasicrystals. 

In the systems investigated in the previous section that are obtained from a liquid, we observe similar interrupted lines of particles which do not find together, at least not within the simulation time. Thus, the fast decay of $g(\bi{r})$ seems to be due to phasons which are not completely relaxed compared to the perfect quasicrystal. 

We were not able to determine beyond doubt whether the equilibrium structure of the quasicrystal should contain the phasonic excitations as observed for the systems started from the liquid or not as seen for systems started from a perfect decagonal order. The excitation or the relaxation of correlated phasonic flips occurs on timescales that are far too long to be studied with Monte Carlo simulations. While we can study the dynamics of local phasonic flips over long times in much smaller systems, we cannot explore the phasonic excitations that matter for the long-distance behavior of the correlation functions and that correspond to rearrangements correlated over long distances which are only accessible with large simulation boxes. Simulations where exclusively phasonic flips can occur indicate that phasonic flips are part of thermal equilibrium \cite{Hielscher2} and the structure can still be dislocation free or at least usually no isolated dislocations occur. 
 
Furthermore, the logarithmic decay of the correlations of phasonic displacements \cite{Kalugin,Henley} prohibits a perfect long-range phasonic order. As a consequence it seems possible that the fast decay of the positional order is an equilibrium feature of two-dimensional quasicrystals that can only be avoided if the quasicrystal is stuck in a metastable state and the equilibrium phasonic excitations take too long to occur.

\section{\label{sec:conclusions}Conclusions}

By using event-chain Monte Carlo simulations, we have examined the phase diagram of two-dimensional systems of particles that interact according to a Lennard-Jones--Gauss potential. While in previous works, the event-chain Monte Carlo algorithm was applied to particles with purely repulsive interactions \cite{Kapfer}, our potential possesses two minima providing two incommensurate length scales as present in decagonal quasicrystals. The analysis of the observed configurations based on their orientational order revealed gas, liquid and quasicrystalline phases as well as coexistence regions.

We observed a first-order transition between quasicrystal and liquid. We did not find any pentahedratic phase loke suggested within the KTHNY theory for quasicrystals \cite{De}. While the orientational order of the solid is long-range, the positional order (suggested by the coherent correlation function) is either quasi long-range as predicted by the Mermin-Wagner theorem \cite{Mermin,Mermin2} when started from a perfect decagonal structure, or short-range when the initial structure is a liquid. We found that excited phasonic degrees of freedom can be responsible for the faster decay of the correlation functions. I.e., since $g(\bi{r})$ is affected by phasons, it is not fundamental for the determination of order in quasiperiodic structures.

Our results demonstrate that intrinsic quasicrystals obtained by special pair interactions \cite{Denton,Barkan,Barkan2,Achim,Dotera,Savitz} can possess phasonic excitations that lead to a fast decay of the positional correlation function. Note that the growth of quasicrystals \cite{Achim,Schmiedeberg3,Martinsons2,Gemeinhardt} is usually accompanied by single or correlated phasonic flips. Only close to the triple point perfect quasicrystals without phasonic flips may grow \cite{Achim}. Similarly, tilings obtained from construction rules can be excitation-free \cite{Elser3} or dominated by phasonic flips \cite{Henley}. It might also be interesting to study the positional order in substrate-induced quasicrystals as obtained in experiments \cite{Foerster,Foerster2}, simulations \cite{Sandbrink2} or theoretical studies \cite{Neuhaus,Neuhaus2}. 

Even though our simulations did not reveal a pentahedratic phase we cannot rule out the possibility to find a melting transition according to the KTHNY theory with an intermediate phase in systems with other quasicrystalline symmetries or structures induced by different pair potentials \cite{Denton,Barkan,Barkan2,Achim,Dotera,Savitz}. Furthermore, one might determine the phase transition with different methods like density functional theory or phase field crystal models for quasicrystals \cite{Archer,Archer2,Subramanian,Barkan} or for anisotropic particles that support quasicrystalline order \cite{vanderLinden,Reinhardt,Gemeinhardt,Gemeinhardt2}.

\ack 
The project was supported by the Deutsche
Forschungsgemeinschaft (DFG) within the Emmy Noether program (Grant
No. Schm 2657/2).

\end{document}